\documentclass{aa} 
\usepackage{amsmath} 
\usepackage{graphicx}
\usepackage{natbib}
\bibpunct{(}{)}{;}{a}{}{,}
\usepackage{txfonts}
\usepackage{lscape}     
\usepackage[T1]{fontenc}
\newcommand{\cmd}{\,cm$^{-2}$}   
\newcommand{\kms}{\,km\,s$^{-1}$}

\newcommand{\lo}{\,$L_{\sun}$}
\begin{document}
%
%
\title{Wind asymmetry imprint in the UV light curves\\ 
of the symbiotic binary SY Mus\footnote{Table 1 is only available online.}}
%
\titlerunning{The asymmetry of SY Mus light curves}
%
%
\author{N.~Shagatova \and A.~Skopal}
\institute{Astronomical Institute, Slovak Academy of Sciences,
        059~60 Tatransk\'{a} Lomnica, Slovakia}
%
\date{Received / Accepted }

\abstract
{
Light curves (LCs) of some symbiotic stars show a different slope 
of the ascending and descending branch of their minimum profile. 
The origin of this asymmetry is not understood well. 
}
{
We explain this effect in the ultraviolet LCs of 
the symbiotic binary SY~Mus. 
}
{
%
We model the continuum fluxes in the spectra obtained by the 
{{\em International Ultraviolet Explorer}} at $10$ wavelengths, from 
$1280$ to $3080$\,\AA. We consider that the white dwarf radiation 
is attenuated by H$^0$ atoms, H$^-$ ions and free electrons in 
the red giant wind. Variation in the nebular component is 
approximated by a sine wave along the orbit as suggested by 
spectral energy distribution models. The model includes asymmetric wind velocity distribution 
and the corresponding ionization structure of the binary. 
}
{
%
We determined distribution of the H$^0$ and H$^+$, as well as upper 
limits of H$^-$ and H$^0$ column densities in the neutral and 
ionized region at the selected wavelengths as functions of the 
orbital phase. 
Corresponding models of the LCs match well the observed continuum 
fluxes. In this way, we suggested the main UV continuum 
absorbing (scattering) processes 
in the circumbinary environment of S-type symbiotic stars. 
}
{
%
The asymmetric profile of the ultraviolet LCs of SY~Mus is caused 
by the asymmetric distribution of the circumstellar matter at the 
near-orbital-plane area. 
}
\keywords{binaries: symbiotic --
          scattering --
          stars: winds, outflows --
          stars: individual: SY~Mus
         }

\maketitle
\section{Introduction}
\label{int}

Asymmetric light curves (LCs) of some interacting binaries along 
the orbital phase can be caused by a variety of geometrical and 
physical effects. For example, by the presence of cool or hot 
spots/areas \citep[e.g.][]{bn80,be84,pu07,sa09,yu10,pr11}, 
an asymmetric wind distribution in symbiotic binaries 
\citep[e.g.][]{du99}, a slopping accretion column \citep{an86}, 
pulsation of binary components 
\citep[][ and references therein]{ka12,po61}, a Coriolis effect 
\citep[e.g.][]{zl90}, or an eccentric orbit \citep{el80}. 
In the X-ray binaries, the asymmetry of the ascending and descending 
parts of LCs can originate from a relativistic Doppler effect 
\citep{wa05}. 

The asymmetry of ingress and egress branch of the LCs minimum was 
also indicated for some symbiotic systems 
\citep[e.g.][]{du99,ko02,sk02,ss+12,wi10}. These widest interacting 
binaries comprise usually a red giant (RG) as the donor star and 
a white dwarf (WD) as the accretor. The binary components interact 
via the wind mass transfer from the RG to the WD 
\citep[e.g.][]{bo67,ms99}. 

Using a semi-quantitative analysis, \cite{fl90} suggested 
that a reflection effect can account for the wave-like orbitally 
related variations in the LCs of three symbiotic systems. 
Analysing long-term photometric observations of a group of classical 
symbiotic stars, \cite{sk98} revealed apparent changes of 
their orbital period as a consequence of variable, orbitally 
related nebular emission. In addition, \cite{sk01} found that 
the observed amplitudes of the LCs are far larger than those 
calculated within a model of the reflection effect. 
%
%
For a supersoft X-ray symbiotic binary SMC3, the asymmetry of LCs 
in the V and I filter and the X-rays was modelled by \cite{ka13}. 
They assumed a spiral tail of neutral hydrogen at the orbital plane 
produced by the giant almost filling in its Roche lobe. 

Using a Monte Carlo simulations of the Rayleigh scattering 
effects in symbiotic stars, \cite{sc95} showed that the eclipse 
width and depth in ultraviolet (UV) LC profiles depend mainly 
on the extension of the H$^0$ region within the model of 
\cite{se84} (hereafter STB), wavelength and contribution 
of the scattered light.

SY~Mus is a quiet eclipsing symbiotic binary, which shows an 
asymmetry in its UV LCs. \cite{du99} demonstrated 
this effect for the continuum fluxes at 1380\,\AA. They found 
that the ascending branch of the LC minimum is less steep 
than the descending one, and shifted more from the position 
of the spectroscopic conjunction. Further, measuring the neutral
hydrogen column densities around the eclipse of the hot component, 
they revealed the asymmetric wind density distribution around 
the giant and suggested this effect to be responsible for 
the asymmetric UV LC around the eclipse. Recently, \cite{ss16} 
derived the velocity profiles of the wind in symbiotic systems 
SY~Mus and EG~And and they indicated that the wind from the giant
 is focused at the 
orbital plane. 

In this work we prove that the observed asymmetry in the UV LC 
profile is primarily caused by the asymmetrical 
displacement of the circumbinary matter at the orbital-plane-area 
with respect to the binary axis. 
In Sect.~2, we present our dataset and in Sect.~3 we introduce 
our model. 
Discussion of results and conclusion  are included in 
Sects.~4 and 5. 

\section{Observed continuum fluxes of SY Mus}
\label{obs}

We used 44 SWP and 39 LWP/LWR International Ultraviolet Explorer (IUE)
spectra of SY Mus to measure the continuum fluxes at 10 
wavelengths from $1280$ to $3080$\,\AA\, (Table~1, available online). 
The spectra were dereddened with $E_{\rm B-V} = 0.35$\,mag 
\citep[][]{sk05}. According to \cite{du99} we used the ephemeris 
for the time of the inferior conjunction of the giant as 
\begin{equation}
  JD_{\rm spec. conj.} = 2\,450\,176 + 625\times E .
\label{eq:eph}
\end{equation}
%
The resulting LCs show an asymmetry of their descending and 
ascending branches and an offset of the minima position with 
respect to the time of the inferior conjunction of the giant. 
These effects weaken towards the longer wavelengths 
(Fig.~\ref{10lc}).
%
%
\section{The model}
\label{mod}
 
\subsection{Components of the continuum radiation}
\label{rad}

During quiescent phases, the continuum radiation of S-type (stellar) 
symbiotic binaries consists of three basic components. 
Radiation from the WD dominates the far-UV
range ($\lambda\lesssim 1800$\,\AA), whereas the ionized part of 
the wind, i.e. the symbiotic nebula, represents the main 
contribution within the near-UV and optical 
(Fig.~\ref{neb}, bottom). The radiation from 
the RG dominates the spectrum from around $VRI$ passbands to 
longer wavelengths, depending on its spectral type \citep[e.g.][]{sk05}. 
Therefore, 
we assume the WD and nebula as the only sources of radiation 
in analysing the UV continuum (120 -- 330\,nm) of SY~Mus. 
%
%
To model the observed orbitally-related flux variation, we use 
the following assumptions. 
\begin{itemize}
\item
The hot component radiates as a black-body at the temperature 
$T_{\rm h}=105\,000$\,K \citep[][]{mu91}. The radiation is 
attenuated by the wind material along the line of sight as 
a function of the orbital phase $\varphi$, and can be expressed as, 
\begin{equation}
   F^{\rm h}_{\lambda}(\varphi) =\pi B_\lambda(T_{\rm h})
               e^{-\tau_{\lambda}(\varphi)},
\label{Fhot}
\end{equation}
   where $B_\lambda(T_{\rm h})$ is Planck function 
and $\tau_{\lambda}(\varphi)$ is the total optical depth along 
the line of sight. 
\item 
The radiation from the nebula is for the sake of simplicity 
approximated by a sine wave as, 
\begin{equation}
   F^{\rm n}_{\lambda}(\varphi) =\alpha_{\lambda} 
   \sin [2\pi (\varphi-0.25)]+\beta_{\lambda},
\label{Fneb}
\end{equation}
   where $\alpha_\lambda$ and $\beta_\lambda$ are model parameters.
This assumption is based on the fact that the nebular continuum 
varies with the orbital phase \citep[e.g.][]{fc88} and is responsible 
for the wave-like orbitally-related variation observed in the LCs of 
symbiotic stars during quiescence \citep[][]{sk01}. Their profile 
along the orbit can be compared to a sine function 
\citep[for SY~Mus, see Fig.~2 of][]{pe95,sk09}.
\end{itemize}
The model continuum is then given by their superposition, i.e., 
\begin{equation}
   F_{\lambda}(\varphi) = F^{\rm h}_{\lambda}(\varphi)+
   F^{\rm n}_{\lambda}(\varphi).
\label{Ftot}
\end{equation}

\subsection{Attenuation in the stellar continuum}
\label{att}

In this section, we discuss individual components of the total 
optical depth, $\tau_\lambda(\varphi)$, that attenuates the WD continuum 
radiation on the line of sight. With respect to the ionization structure 
of the hydrogen 
in the binary, we consider attenuation processes in the neutral and 
ionized part of the giant's wind. Therefore, $\tau_\lambda(\varphi)$ is 
a sum of the optical depth of the neutral region, 
$\tau_\lambda^0(\varphi)$, and that of the ionized region, 
$\tau_\lambda^+(\varphi)$, i.e.,
\begin{equation}
  \tau_\lambda(\varphi)=\tau_\lambda^0(\varphi)+
  \tau_\lambda^+(\varphi). 
\label{tautot}
\end{equation}
In a dense H$^0$ zone 
of symbiotic binaries, a strong attenuation of the continuum 
around the Ly-$\alpha$ line is caused by Rayleigh scattering 
\citep[][]{nu89}. Further, we consider the scattering
on negative 
hydrogen ion, H$^-$, that attenuates the stellar radiation in cool 
atmospheres \citep[e.g.][]{gr05}. Thus, we can write, 
\begin{equation}
  \tau_\lambda^0(\varphi)=\sigma_{\rm Ray}(\lambda)n_{\rm H^0}(\varphi)+
  \kappa_{\rm H^-}(\lambda)n_{\rm H^-}(\varphi),
\label{tauN}
\end{equation}
where $\sigma_{\rm Ray}(\lambda)$ is the cross-section of Rayleigh 
scattering \citep[Eq.~5 and Fig.~2 of][]{nu89}, $n_{\rm H^0}(\varphi)$ and 
$n_{\rm H^-}(\varphi)$ is the function for the column density of the neutral 
hydrogen (Sect.~\ref{dens}) and negative hydrogen ion, respectively.
The H$^-$ bound-free 
absorption coefficients, $\kappa_{\rm H^-}(\lambda)$, are given 
by \cite{ge62}. We used $\kappa_{\rm H^-}(\lambda)$ values 
from the "velocity"-curve plotted in his Fig.~6. Corresponding $\kappa_{\rm H^-}(\lambda)$ curve reaches its maximum 
at $\approx 8500$\,\AA. 

\begin{figure}
\centering
\begin{center}
\resizebox{\hsize}{!}{\includegraphics[angle=0]{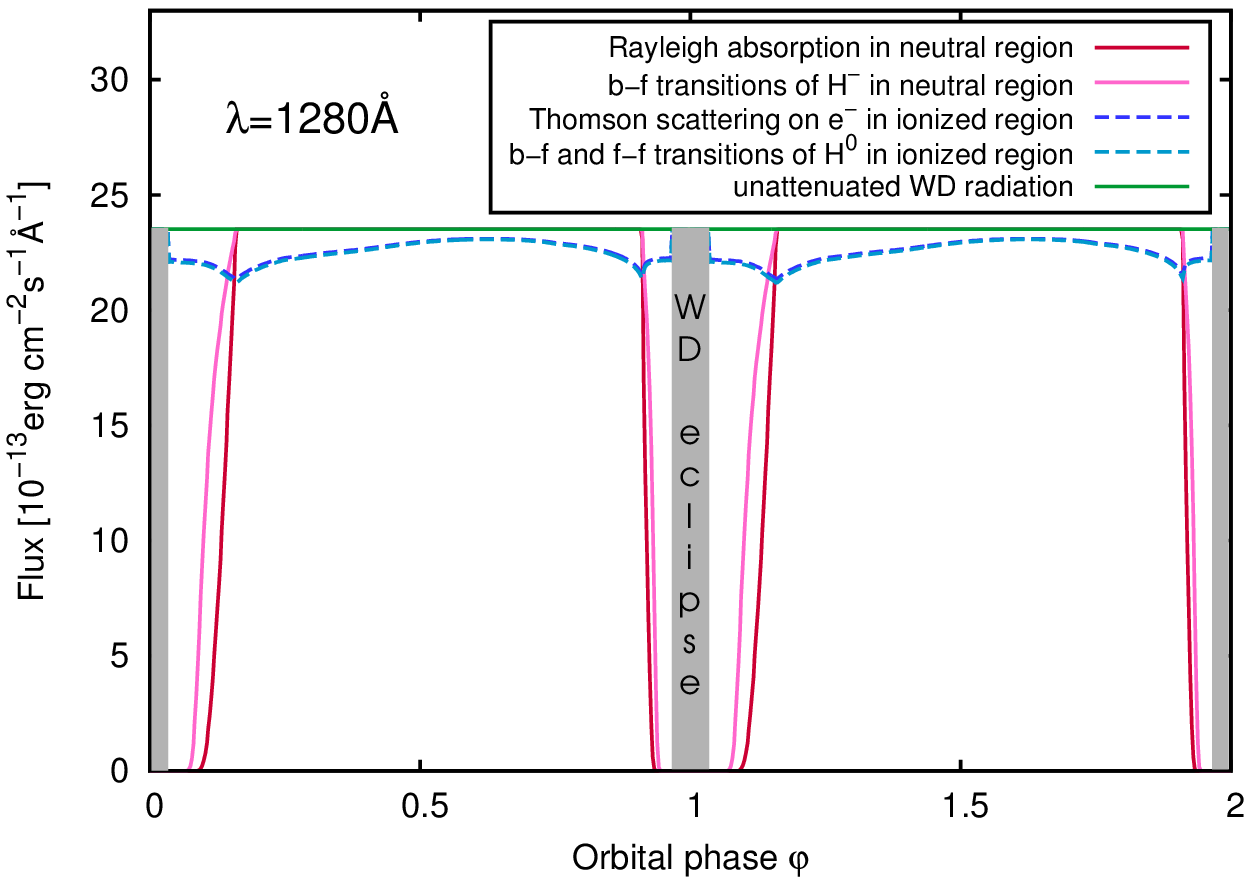}}
\resizebox{\hsize}{!}{\includegraphics[angle=0]{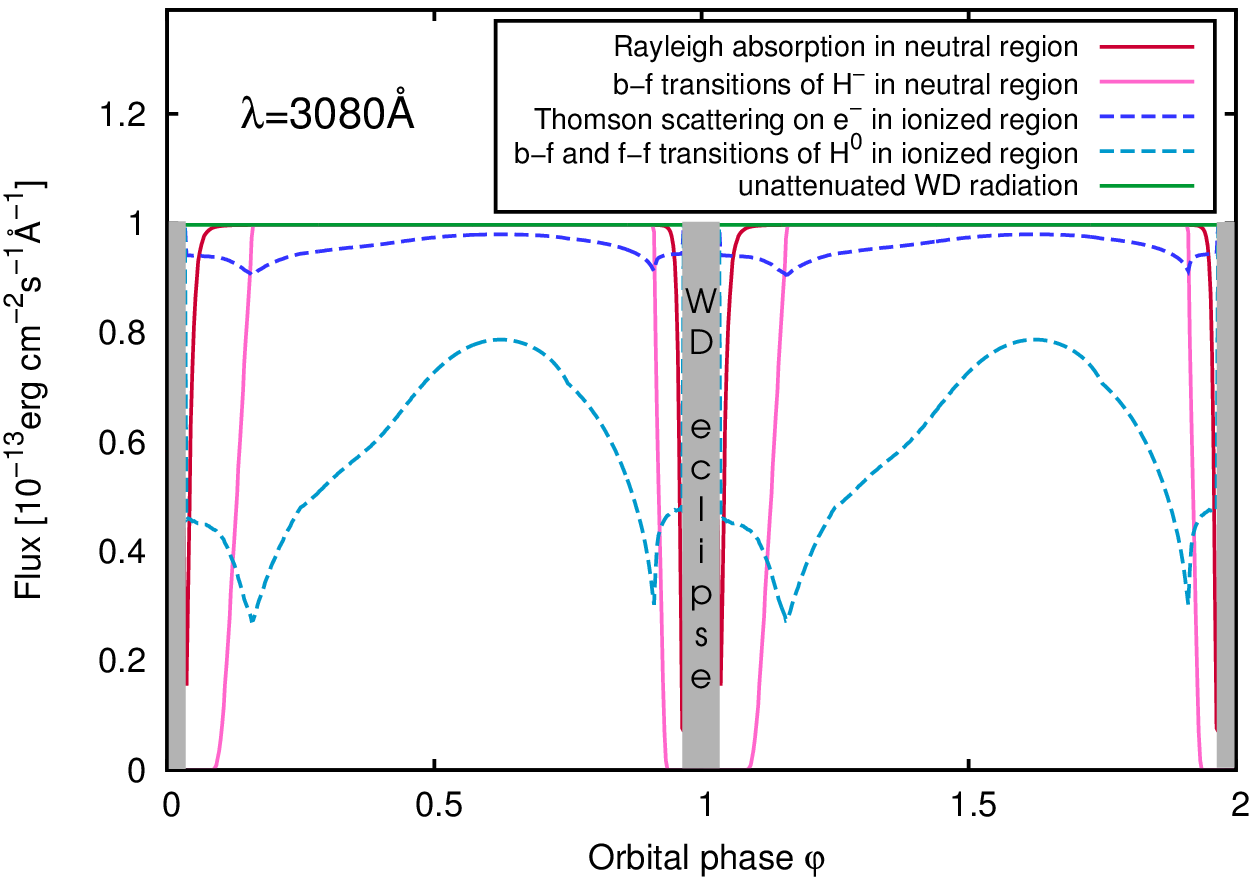}}
\end{center}
\caption[]{
Comparison of the effect of individual considered attenuation 
processes on the WD flux (Sect.~\ref{att}) at shortest (top) and longest 
(bottom) wavelength of the set. Corresponding column densities of H$^0$ 
in neutral region and H$^+$ in ionized region are given by the column 
density model in Fig.~\ref{nHvsetky} and of H$^{-}$ in neutral region 
and H$^0$ in ionized region by the LC modelling (Sect.~\ref{cur}).
Two minima of radiation out of the eclipse for attenuation processes 
in ionized region (dashed lines) correspond to the orbital phase with 
line of sight entering/exiting the neutral region.
         }
\label{abscomp}
\end{figure}

We assume that the contribution to the H$^-$ 
absorption coefficient due to free-free transitions is negligible, 
considering the absorption coefficient values given in
\cite{ge62}, \cite{df66} and \cite{gr05}. The dependence of the H$^-$ 
column density on the orbital phase, $n_{\rm H^-}(\varphi)$, 
remains a variable in the LC modelling through 
the neutral region. The values of $\sigma_{\rm Ray}(\lambda)$ 
and $\kappa_{\rm H^-}(\lambda)$ for selected wavelengths are summarized in 
Table~\ref{abscoef}. The Rayleigh scattering is the dominant 
source of 
opacity in the neutral region at $\lambda=1280$\,\AA, however, its 
effect on the WD radiation is comparable with b-f transitions of 
H$^-$ ion at $\lambda=1340$\,\AA. At longer wavelengths, these 
transitions have the dominant attenuating effect in the neutral region
(see Fig.~\ref{abscomp}).

\begin{figure}
\centering
\begin{center}
\resizebox{\hsize}{!}{\includegraphics[angle=0]{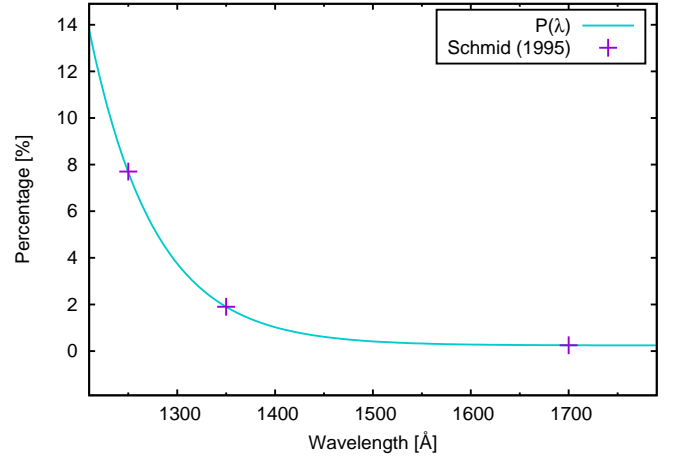}}
\end{center}
\caption[]{
Percentage $P(\lambda)$ of the Rayleigh scattered photons in 
direction of the line of sight fitted to the data from the model S$_{B3}$ 
of \cite{sc95}.
}
\label{Plambda}
\end{figure}

In Eq.~\eqref{tauN}, Rayleigh scattering is treated as an 
absorption process. However, the Rayleigh scattered photons can contribute 
to the line-of-sight continuum 
radiation. \cite{sc95} used Monte Carlo simulations to determine the 
percentage of these photons in the total flux. In his model S$_{B3}$, 
corresponding to a typical symbiotic system, with model parameters 
closest to that of SY Mus, he obtained values of $7.7\%$ at 
$\lambda=1250$\,\AA, $1.9\%$ at $\lambda=1350$\,\AA\, and $0.25\%$ at 
$\lambda=1700$\,\AA, as seen at the orbital phase $\varphi=0.25$. To 
estimate this effect in our LC modelling , we fitted his values by a function 
\begin{equation}
 P(\lambda)=a_0+e^{a_1(a_2-\lambda)},
\label{eqPlambda}
\end{equation}
which represents the percentage of flux from Rayleigh scattered photons 
in the total modelled flux. This function for resulting values of 
parameters  $a_0=0.241$, $a_1=0.015$\,\AA$^{-1}$ and 
$a_2=1383,66$\,\AA\, is depicted in Fig.~\ref{Plambda} and evaluated in
Table~\ref{abscoef}. The contribution from the Rayleigh scattered photons 
can vary with orbital phase. However, according to Fig. 8 of \cite{sc95}, 
these variations are not significant for the model S$_{B3}$ at 
$\lambda\geq 1280$\,\AA. Therefore, we neglect them in our modelling. 
Accordingly, we scale the total flux \eqref{Ftot} as 
$\frac{100}{100-P(\lambda)}F_\lambda(\varphi)$ for $\lambda\leq 1700$\,\AA\, 
and set $P(\lambda)\equiv 0$ for $\lambda> 1700$\,\AA.

In the predominantly ionized region, the high photon flux from 
the hot components 
\citep[$L_{\rm H}\approx 10^{2} - 10^{4}$\lo,][]{mu91,gr97,sk05} produces 
free electrons by ionizing the neutral hydrogen in the RG's wind.
Column densities of the free electrons during quiescent phases of symbiotic 
stars, $n^+_{\rm e}\approx 10^{22}-10^{23}$\cmd\, were determined by 
\cite{ses12} assuming that the Thomson scattering is responsible for the 
broad wings of the strongest emission lines in the far-UV. Their values of 
$n^+_{\rm e}$ 
are consistent with our model 
based on H$^0$ column density modelling (see below Eq.~\eqref{eqne} and 
Fig.~\ref{nHvsetky}). Therefore, we suppose that 
the free electrons can be relevant for the attenuation of the 
stellar UV continuum
by Thomson scattering. Further, 
the bound-free and free-free transitions on neutral hydrogen 
can be effective at a high electron temperature of nebula 
 \citep[$T_{\rm e}\approx 18500$\,K,][]{sk05}. 

Under these conditions, the 
optical depth through the ionized part of the wind can be written as,
\begin{equation}
  \tau_\lambda^+(\varphi)=\sigma_{\rm e}n^+_{\rm e}(\varphi)+
  \sigma_{\rm H^0}(\lambda,\rm T_e)n^+_{\rm H^0}(\varphi),
\label{tauI}
\end{equation}
where $\sigma_{\rm e}$ is the Thomson scattering cross-section, 
$\sigma_{\rm H^0}(\lambda,\rm T_e)$ the total cross-section of neutral 
hydrogen and $n^+_{\rm H^0}(\varphi)$ is the column density of H$^0$ in 
the symbiotic nebula. 
We set the free electron column density 
\begin{equation}
n^+_{\rm e}(\varphi)=
1.2\,n^+_{\rm p}(\varphi),  
\label{eqne}
\end{equation}
where $n^+_{\rm p}(\varphi)$ is 
the column density of protons in the ionized area. 
The cross-section 
$\sigma_{\rm H^0}(\lambda,\rm T_e)$ accounts for both bound-free 
and free-free transitions, i.e., $\sigma_{\rm H^0}(\lambda,\rm T_e)=
f_{\rm se}(\lambda,\rm T_e)(\kappa_{\rm bf}(\rm H^0)+
\kappa_{\rm ff}(\rm H^0))$, where $\kappa_{\rm bf}(\rm H^0)$ 
is the absorption coefficient for bound-free transitions and 
$\kappa_{\rm ff}(\rm H^0)$ for free-free transitions of neutral hydrogen 
atom \citep[e.g. Eq.~(8.8) and (8.10) of ][; for the used values, 
see Table \ref{abscoef}]{gr05}. 
The factor $f_{\rm se}$ accounts for the stimulated emission and 
$n^+_{\rm H^0}(\varphi)$ represents the variable 
in the LC modelling through the ionized wind region. 
In ionized region, the Thomson scattering and transitions of H$^0$ 
have comparable attenuation effect on the WD radiation at $\lambda=1280$\,\AA, 
while the latter becomes dominant source of opacity with increasing wavelength 
(Fig.~\ref{abscomp}). In Sect.~\ref{dis}, we discuss the effect of other ions to 
the continuum attenuation.

Finally, we note that the description of the above-mentioned 
processes is valid in nebulae under conditions of the local 
thermodynamic equilibrium (LTE). This assumption is supported 
by the fact that the free electrons thermalize in nebulae very 
quickly, because of an extremely short time of their electrostatic 
encounters with respect to any other inelastic scatterings 
\citep[see][]{ba47}. As a result, the energy distribution of free electrons can 
be characterized by a single 
$T_{\rm e}$. Modelling the UV/optical/near-IR spectral energy distribution (SED) during 
quiescent phases supports this assumption 
\citep[see figures and Table~3 of][]{sk05}. However, e.g. 
\cite{pr98} used a non-LTE photoionization code to calculate the spectrum of a red giant wind illuminated by the hot component of a symbiotic binary. Therefore, the determination of H$^0$ column density 
in ionized region can suffer from a systematic error in our simplified approach.

\setcounter{table}{1}
\begin{table}[t!]
   \caption{
Values of the Rayleigh scattering cross-section 
$\sigma_{\rm Ray}(\lambda)$ (in cm$^2$), the percentage of the Rayleigh scattered 
photons in direction of the line of sight  (in \%), H$^-$ 
bound-free absorption coefficient $\kappa_{\rm H^-}(\lambda)$ and total cross-section 
of neutral hydrogen 
$\sigma_{\rm H^0}(\lambda,\rm T_e)$ at $\rm T_e=18500$\,K (both in cm$^2$) for 10 selected wavelengths.
           }
\label{abscoef}
\centering
\begin{tabular}{ccccc}
\hline\hline
$\lambda$\,[\AA] & $\sigma_{\rm Ray}(\lambda)$ & $P(\lambda)$ & $\kappa_{\rm H^-}(\lambda)$ 
    & $\sigma_{\rm H^0}(\lambda,\rm T_e$)   \\
\hline
1280 & $1.096\times 10^{-23}$ & 5.0 & $4.56\times 10^{-18}$ & $5.755\times 10^{-21}$ \\
1340 & $2.984\times 10^{-24}$ & 2.2 & $4.85\times 10^{-18}$ & $6.580\times 10^{-21}$ \\
1450 & $8.290\times 10^{-25}$ & 0.6 & $5.51\times 10^{-18}$ & $8.263\times 10^{-21}$ \\
1520 & $4.801\times 10^{-25}$ & 0.4 & $5.81\times 10^{-18}$ & $9.470\times 10^{-21}$ \\
1600 & $2.911\times 10^{-25}$ & 0.3 & $6.20\times 10^{-18}$ & $1.100\times 10^{-20}$ \\
1800 & $1.140\times 10^{-25}$ & 0.0 & $7.31\times 10^{-18}$ & $1.545\times 10^{-20}$ \\
1950 & $6.663\times 10^{-26}$ & 0.0 & $8.07\times 10^{-18}$ & $1.938\times 10^{-20}$ \\
2400 & $2.019\times 10^{-26}$ & 0.0 & $1.10\times 10^{-17}$ & $3.521\times 10^{-20}$ \\
2700 & $1.107\times 10^{-26}$ & 0.0 & $1.31\times 10^{-17}$ & $4.936\times 10^{-20}$ \\
3080 & $5.874\times 10^{-27}$ & 0.0 & $1.58\times 10^{-17}$ & $7.176\times 10^{-20}$ \\
\hline
\end{tabular}
\end{table}
%

\subsection{Column densities distribution}
\label{dens}

To calculate the total optical depth (Eq.~\eqref{tautot}) along 
the line of sight to the WD, we need to determine functions 
$n_{\rm H^0}(\varphi)$ and $n^+_{\rm p}(\varphi)$. According 
to STB, we determine the ionization 
boundary by solving the parametric equation \citep[see][]{nv87}, 
\begin{equation}
   X^{\rm H+} = f(u,\vartheta). 
\label{ionhr}
\end{equation}
It defines the boundary between neutral and ionized hydrogen 
at the plane of observations by polar coordinates ($u,\vartheta$) 
with the origin at the hot star. The ionization parameter 
$X^{\rm H+}$ is expressed as 
\begin{equation}
   X^{\rm H+} = \frac{4\pi(\mu m_{\rm H})^2}
                 {\alpha_{\rm B}({\rm H},T_{\rm e})} p L_{\rm ph}
            \left(\frac{v_\infty}{\dot M}\right)^2, 
\label{XH+}
\end{equation}
where $\mu$ is the mean molecular weight, $m_{\rm H}$ the mass of the 
hydrogen atom, 
$\alpha_{\rm B}({\rm H},T_{\rm e})$ the total hydrogen 
recombination coefficient for recombinations other than to the 
ground state (Case B), $p$ 
the binary separation, $L_{\rm ph}$ the flux of ionizing photons from 
the WD, $v_\infty$ the terminal velocity of the wind from the giant 
and $\dot{M}$ its mass-loss rate.

Then, we can obtain the column density of neutral hydrogen by 
integrating H$^0$ number density along the line of sight from the 
observer ($-\infty$) to the position of the ionization boundary 
$l_\varphi$, 
\begin{equation}
  n_{\rm H^0}(b) = \displaystyle\frac{a}{2}
          \displaystyle\int\limits_{-\infty}^{l_{\varphi}(b)}
          \displaystyle\frac{dl}{(l^2+b^2)v(\sqrt{l^2+b^2})}.
\label{nHnum}
\end{equation}
where $a=\dot{M}/2\pi\mu m_{\rm H}$, $l$ is the coordinate along 
the line of sight, $b$ is the impact parameter 
\citep[see][]{ss12,ss16} and $v(\sqrt{l^2+b^2})$ is the wind velocity 
distribution (Sect. \ref{vfi}). The impact parameter represents the 
position of the binary through the orbital inclination $i$ and 
phase angle $\varphi$ as 
$b^2=p^2(\cos^2 i + \sin^2 \varphi \sin^2 i)$. 
The value of the integral (\ref{nHnum}) decreases from the inferior 
conjunction of the RG and vanishes at the orbital phases, where 
the line of sight does not cross the H$^0$ zone.

Similarly, we can obtain the column density of protons in the ionized 
area in the form, 
\begin{equation}
  n^+_{\rm p}(b)=\displaystyle\frac{a}{2}
  \displaystyle\int\limits_{-\infty/l_{\varphi}(b)}^{\pm\sqrt{p^2-b^2}}
  \displaystyle\frac{dl}{(l^2+b^2)v(\sqrt{l^2+b^2})},
\label{nH+num}
\end{equation}
where the lower limit of integration is $l_\varphi(b)$ for orbital 
phases, at which the line of sight passes through the neutral area, 
and $-\infty$ outside it. The sign plus in the upper limit of 
integration (i.e. the position of the WD) applies for values 
of the angle between the line of sight and the line connecting 
the binary components, 
$\vartheta\in\langle-\pi/2,\pi/2\rangle$, and the sign minus applies 
for 
$\vartheta\in\langle-\pi,-\pi/2\rangle\cup\langle\pi/2,\pi\rangle$.

For the orbital phases during the eclipse, i.e. for 
$b\leq R_{\rm g}$ (= the radius of the RG), we integrated 
$n_{\rm H^0}(b)$ up to the giant surface, i.e. 
$l=-\sqrt{R_{\rm g}^2-b^2}$, and put $n^+_{\rm p}(b)=0$.
%

\subsection{Velocity profiles}
\label{vfi}

\subsubsection{Components derived from observations}
\label{vevin}

The functions for the column densities of neutral and ionized hydrogen 
(Eq.~\eqref{nHnum} and \eqref{nH+num}) are determined by the wind velocity 
distribution.
We adopt the velocity profiles from \cite{ss16}, derived by 
modelling the H$^0$ column densities, measured from the Rayleigh attenuation 
around the Ly-$\alpha$ line during ingress and egress orbital phases, using 
the inversion method of \cite{kn93} and the ionization structure model STB.
We use their egress model M and ingress model O,
denoting
them here as $v_{\rm e}(r)$ and $v_{\rm in}(r)$:
\begin{equation}
\label{vinr}
v_{\rm in}(r)=\displaystyle\frac{v_{\rm in\,\infty}}
{1+1.81\times 10^4\displaystyle\left( \frac{r}{R_{\rm g}}\right)^{-12} }
\end{equation}
and
\begin{equation}
\label{ver}
v_{\rm e}(r)=\displaystyle\frac{v_{\rm e\,\infty}}
{1+2.94\times 10^3\displaystyle\left( \frac{r}{R_{\rm g}}\right)^{-7} },
\end{equation} 
where $v_{\rm in\,\infty}$ and $v_{\rm e\,\infty}$ are corresponding 
terminal velocities of the wind.
The ingress velocity profile is steeper than the egress profile, thus, 
it indicates an asymmetric wind distribution at the plane of observations. 
For eclipsing binaries, the plane of observations roughly coincides with 
the orbital plane.

For the egress terminal velocity we adopted 
$v_{\rm e\infty}=20$\kms, which is within a typical range of the terminal 
velocities of symbiotic stars \citep[e.g.][]{du99,sc99}. The selected 
value of $v_{\rm e\infty}$ yields $\dot{M}$ of the model, 
using Eq.~(18) of \cite{ss16}. However, this value represents the spherical 
equivalent of the mass-loss rate because of the asymmetry in the wind distribution 
around the RG.

\subsubsection{Model of a complete velocity profile}
\label{vpsi}

As we aim to model the continuum LCs of SY~Mus for the complete range 
of phases, we need to propose a unified model of the velocity profile by 
interconnecting $v_{\rm in}(r)$ and $v_{\rm e}(r)$ components around
the orbital phase $0$, where the observations of column densities are
not available; i.e. for the so-called transition area. We assume
that the velocity profile changes gradually
in a smooth way from $v_{\rm in}(r)$ 
to $v_{\rm e}(r)$ at the plane of observations, i. e. we are looking for the 
transition velocity profile, $v_{\rm trans}(r,\psi)$, where $\psi$ is the 
azimuth angle at the plane of observations in the coordinate system centered 
at the RG. Thus, the $v_{\rm trans}(r,\psi)$ function is defined for 
$\psi\in\langle\psi_{\rm in},\psi_{\rm e}\rangle$, where angles $\psi_{\rm in}$ 
and $\psi_{\rm e}$ bound the transition area.
In determining the $v_{\rm trans}(r,\psi)$ function, we require the following
conditions:
\begin{itemize}
\item the unified function for the velocity profile has to be a smooth 
(C$^1$) function of $\psi$,
\item corresponding functions, $n_{\rm H^0}(\varphi)$, $n^+_{\rm p}(\varphi)$, 
and the position of the ionization boundary, $l_\varphi$, have to be 
smooth (C$^1$) functions of $\varphi$,
\item unified model of $n_{\rm H^0}(\varphi)$ has to fit the observed H$^0$ 
column densities. 
\end{itemize}
The transition velocity profile affects the model density distribution of the wind 
mainly at the area corresponding to the lines of sight during the eclipse.

\begin{figure}
\centering
\begin{center}
\resizebox{\hsize}{!}{\includegraphics[angle=0]{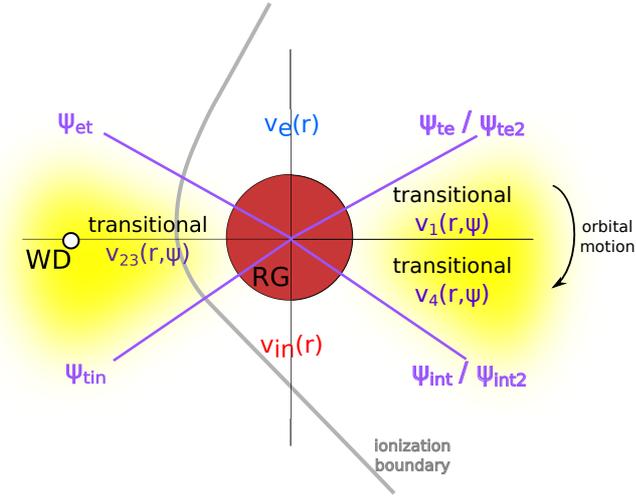}}
\end{center}
\caption[]{Schematic view of the regions of different velocity 
profiles 
as seen perpendicularly to the plane of observations. 
Denotations are explained in Sect.~\ref{vfi}. 
}
\label{4kvPrech}
\end{figure}

Since a linear function of $\psi$ cannot meet the conditions for 
$v_{\rm trans}(r,\psi)$, we assume the transition velocity profile in the form:
\begin{equation}
\label{vtrans}
v_{\rm trans}(r,\psi)=v_{\rm in}(r)\cos^2(C_{\rm in}\psi + D_{\rm in})+
v_{\rm e}(r)\cos^2(C_{\rm e}\psi + D_{\rm e}),
\end{equation}
where $C_{\rm in}$, $D_{\rm in}$, $C_{\rm e}$ and $D_{\rm e}$ are parameters. 
We divided the complete range of $\psi$, from $0$ to 
$2\pi$, into 4 areas with different velocity profiles, as depicted in 
Fig.~\ref{4kvPrech}. In front of the RG along the orbital motion, there is an 
area, where the $v_{\rm e}(r)$ profile applies. At the opposite side, the 
$v_{\rm in}(r)$ profile is implemented. Finally, there are two more zones in 
between, where the transitional velocity profile applies.
From the computational point of view, the $v_{\rm trans}(r,\psi)$ profile is 
defined by three functions: $v_{1}(r)$, $v_{23}(r)$ and $v_{4}(r)$. The 
interconnection of $v_{\rm in}(r)$ and $v_{\rm e}(r)$ from the quadrant II to 
III of the values of $\psi$
is given by $v_{23}(r)$, the interconnection between the quadrant I and IV is 
given by two functions, $v_{1}(r)$ and 
$v_{4}(r)$, because of the discontinuity in values of $\psi$ around $\psi=0=2\pi$. 
The axisymmetric boundaries of $v_{\rm e}(r)$ and $v_{\rm in}(r)$ 
regions are given by relations between limiting values of $\psi$ 
(see Fig.~\ref{4kvPrech}):
%
\begin{center}
  $\hspace{-1.2cm}\psi_{\rm te} \in \langle 0,\pi/2\rangle, \hspace{1.95cm} 
   \psi_{\rm int} \in \langle -\pi/2,0\rangle$,
\end{center}
\vspace{-0.45cm}
\begin{center}
  $\hspace{-1.45cm}\psi_{\rm et} = \pi - \psi_{\rm te}, \hspace{2.05cm} 
   \psi_{\rm tin} = \pi - \psi_{\rm int}$,
\end{center}
\vspace{-0.2cm}
\begin{equation}
\label{psiRovn}
 \hspace{-1.2cm}\psi_{\rm te2} = 2\pi + \psi_{\rm te},  \vspace{3.0mm} 
 \hspace{1.75cm} 
          \psi_{\rm int2} = 2\pi + \psi_{\rm int},
\end{equation}
where the subscript "te" denotes the boundary from transitional to egress 
velocity profile area and "et" the boundary for the transition in the opposite 
direction, in the direction of increasing values of $\psi$.
Similarly, the subscript "int" denotes the boundary from ingress to transitional 
zone and "tin" from the transitional to ingress zone.
Then, from the conditions of smoothness of the united velocity profile 
(i.e. the first derivative of the corresponding function has to be 
continuous for C$^1$ function), 
we obtain the formulae for velocity profiles defining $v_{\rm trans}(r,\psi)$:
\begin{equation}
\label{vtrans1}
v_{1}(r,\psi)=v_{\rm in}(r)\cos^2[C(\psi-\psi_{\rm int})]+
v_{\rm e}(r)\cos^2[C(\psi-\psi_{\rm te})],
\end{equation}
\begin{equation}
\label{vtrans23}
v_{23}(r,\psi)=v_{\rm in}(r)\cos^2[C(\psi_{\rm tin}-\psi)]+
v_{\rm e}(r)\cos^2[C(\psi-\psi_{\rm et})],
\end{equation}
\begin{equation}
\label{vtrans4}
v_{4}(r,\psi)=v_{\rm in}(r)\cos^2[C(\psi_{\rm int2}-\psi)]+
v_{\rm e}(r)\cos^2[C(\psi-\psi_{\rm te2})],
\end{equation}
where $C=\dfrac{\pi}{2(\psi_{\rm int}-\psi_{\rm te})}$, 
and corresponding transitional terminal velocities ($r\rightarrow\infty$) 
are:
\begin{equation}
\label{vnektrans1}
v_{1\,\infty}(\psi)=v_{\rm in\,\infty}\cos^2[C(\psi-\psi_{\rm int})]+
v_{\rm e\,\infty}\cos^2[C(\psi-\psi_{\rm te})],
\end{equation}
\begin{equation}
\label{vnektrans23}
v_{23\,\infty}(\psi)=v_{\rm in\,\infty}\cos^2[C(\psi_{\rm tin}-\psi)]+
v_{\rm e\,\infty}\cos^2[C(\psi-\psi_{\rm et})],
\end{equation}
\begin{equation}
\label{vnektrans4}
v_{4\,\infty}(\psi)=v_{\rm in\,\infty}\cos^2[C(\psi_{\rm int2}-\psi)]+
v_{\rm e\,\infty}\cos^2[C(\psi-\psi_{\rm te2})].
\end{equation}
Equations \eqref{vtrans1} -- \eqref{vtrans4} and also powers of their right 
sides to the $-1$ and $-2$, which are present in formulae for the position 
of the ionization boundary (Eq.~(8) of \cite{nv87}) and column density 
(Eq.~\eqref{nHnum} and \eqref{nH+num}) fulfill the conditions of smoothness.
The last condition for $v_{\rm trans}(r,\psi)$, i. e. the correctness of 
fits to the observed H$^0$ column densities, is linked with the selection of 
the values of $\psi_{\rm te}$ and $\psi_{\rm int}$ (see Sect.~\ref{ion} 
and \ref{cur}).


\subsection{Ionization parameter}
\label{ion}

To determine the position of the ionization boundary $l_\varphi$, 
values of the ionization parameter $X^{\rm H+}$ are needed. Similarly 
as in the previous section, we adopt them from the column density models M 
and O of \cite{ss16}, i. e., we have $X^{\rm H+}_{\rm in} = 16$ before the 
eclipse and $X^{\rm H+}_{\rm e} = 2.5$ after it. Therefore, 
using Eqs.~\eqref{XH+} and \eqref{vnektrans1} -- \eqref{vnektrans4}, we obtain 
the formulae for the transitional ionization parameters:
\begin{eqnarray}
\label{Xtrans1}
  X^{\rm H+}_{1}(\psi) & = & X^{\rm H+}_{\rm in}\cos^4[C(\psi-\psi_{\rm int})]
   \nonumber \\
  & & +2\sqrt{X^{\rm H+}_{\rm in}X^{\rm H+}_{\rm e}}\cos^2[C(\psi-\psi_{\rm int})]
  \cos^2[C(\psi-                   \psi_{\rm te})] \\
  & & + X^{\rm H+}_{\rm e}\cos^4[C(\psi-\psi_{\rm te})], \nonumber
\end{eqnarray}
\begin{eqnarray}
\label{Xtrans23}
  X^{\rm H+}_{23}(\psi) & = & X^{\rm H+}_{\rm in}\cos^4[C(\psi_{\rm tin}-\psi)]
   \nonumber \\
  & & +2\sqrt{X^{\rm H+}_{\rm in}X^{\rm H+}_{\rm e}}\cos^2[C(\psi_{\rm tin}-\psi)
  ]\cos^2[C(\psi-\psi_{\rm et})] \\
  & & + X^{\rm H+}_{\rm e}\cos^4[C(\psi-\psi_{\rm et})], \nonumber
\end{eqnarray}
\begin{eqnarray}
\label{Xtrans4}
  X^{\rm H+}_{4}(\psi) & = & X^{\rm H+}_{\rm in}\cos^4[C(\psi_{\rm int2}-\psi)]
   \nonumber \\
  & & +2\sqrt{X^{\rm H+}_{\rm in}X^{\rm H+}_{\rm e}}\cos^2[C(\psi_{\rm int2}-\psi)]
  \cos^2[C(\psi-\psi_{\rm te2})] \\
  & & + X^{\rm H+}_{\rm e}\cos^4[C(\psi-\psi_{\rm te2})], \nonumber
\end{eqnarray}
with the subscript notation equivalent to the previous section.

Further, under the assumption of spherically symmetric mass flux at 
$r\rightarrow\infty$, 
we can put into equality the expression for $\dot{M}$ from Eq.~\eqref{XH+} 
for ingress values of variables with that from the same equation 
for egress values of variables to obtain a relation,
\begin{equation}
\label{Xvpom}
\dfrac{X^{\rm H+}_{\rm in}}{X^{\rm H+}_{\rm e}} = \dfrac{v_{\rm in\,\infty}^2}
{v_{\rm e\,\infty}^2}
\end{equation}
and, then, by evaluating egress and ingress ionization parameters, we obtain
a relation between terminal velocities before and after the eclipse,
\begin{equation}
\label{vinve}
v_{\rm in\,\infty}=2,53v_{\rm e\,\infty}.
\end{equation}
This equation reflects the simple distribution of ionization parameter 
values \eqref{Xtrans1} - \eqref{Xtrans4} at the orbital plane and the fact that 
values of $X^{\rm H+}_{\rm in}$ and $X^{\rm H+}_{\rm e}$ were determined by fitting 
$n^{\rm obs}_{\rm H^0}$ values only at the area close to RG (Fig.\ref{nHvsetky}). Since the derivation of Eq.~\eqref{vinve} is based on the equal mass-loss in 
front and behind the RG with respect to its orbital motion, the difference in 
corresponding hydrogen densities in radial directions results from the different velocity 
profiles 
\eqref{vinr} and \eqref{ver}.

The selection of values of $\psi_{\rm int}$ and $\psi_{\rm te}$ has to satisfy 
several conditions. First, it has to
preserve the correctness of the M and O column density models of \cite{ss16}. 
At the same time, the range of the angle $\psi$, where the transitional 
velocity profile and ionization parameter apply, has to be sufficiently wide to 
prevent the corresponding interconnection between the ingress and egress ionization 
boundary 
to be too sharp.

\section{The results}
\label{cur}

\begin{figure}
\centering
\begin{center}
\resizebox{8cm}{!}{\includegraphics[angle=0]{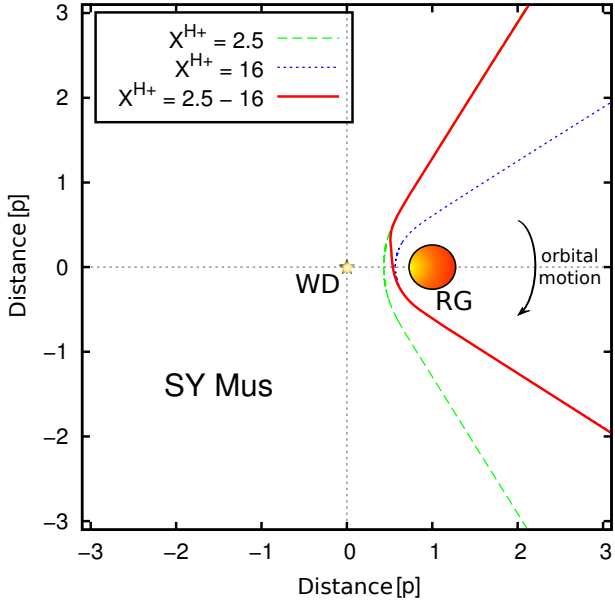}}
\end{center}
\caption[]{Resulting ionization boundary for the united wind velocity model 
given by Eqs.~\eqref{vinr}, \eqref{ver} and \eqref{vtrans1} -- \eqref{vinve}, as seen 
perpendicularly to the 
orbital plane (solid line). The corresponding limiting values of $\psi$ are 
$\psi_{\rm int}= -0.9$\,rad 
and $\psi_{\rm te}= 0.8$\,rad. For comparison, dotted line represents the ionization 
boundary for $X^{\rm H+}=X_{\rm in}^{\rm H+}$ and 
$v(r)=v_{\rm in}(r)$, whereas dashed line for $X^{\rm H+}=X_{\rm e}^{\rm H+}$ 
and $v(r)=v_{\rm e}(r)$. 
          }
\label{ionhranSYMus}
\end{figure}

In this section, we use Eqs.~\eqref{Fhot} - \eqref{vinve} to model the 
asymmetric UV continuum LCs obtained from low-resolution IUE spectra (Table~1). Main results may be summarized as follows:

(i)
We determined the ionization structure by solving Eq.  
\eqref{ionhr} with the united velocity profile 
\eqref{vtrans1} - \eqref{vtrans4} 
and the corresponding ionization parameter \eqref{Xtrans1} - \eqref{Xtrans4}. 
The resulting shape of the 
ionization boundary is depicted in Fig.~\ref{ionhranSYMus}.

\begin{figure}
\centering
\begin{center}
\resizebox{\hsize}{!}{\includegraphics[angle=0]{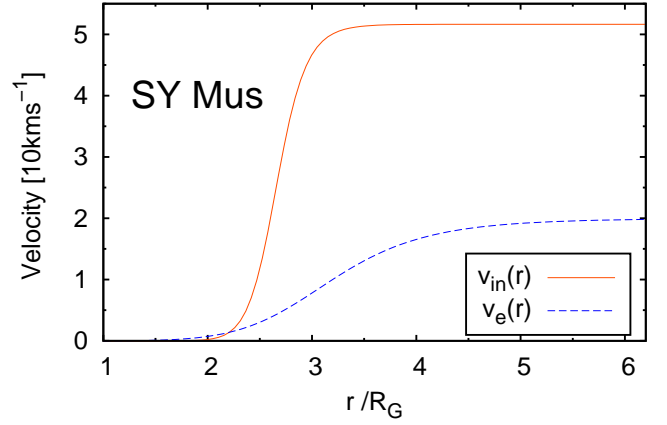}}
%
\end{center}
\caption[]{Ingress and egress velocity profiles of SY Mus from models M 
and O of \cite{ss16} for $v_{\rm e\infty}=20$\kms\,satisfying Eq.~\eqref{vinve}.
          }
\label{vnekIE}
\end{figure}

(ii)
Comparing the wind 
velocity profiles $v_{\rm in}(r)$ and $v_{\rm e}(r)$
(Fig.~\ref{vnekIE}), the ingress wind velocity is lower than the
egress velocity at the distance up
to $\approx 1$\,R$_{\rm g}$ from the RG surface. It implies a higher hydrogen
number density of the wind in front of the RG than behind it, in the direction of the 
orbital motion. On the other hand, at larger distances, the 
wind velocity and thus, the wind number density proportion is opposite. 
The asymmetry of the ionization 
boundary in Fig.~\ref{ionhranSYMus} is in agreement with this 
fact.

%
%

\begin{figure}
\centering
\begin{center}
\resizebox{\hsize}{!}{\includegraphics[angle=0]{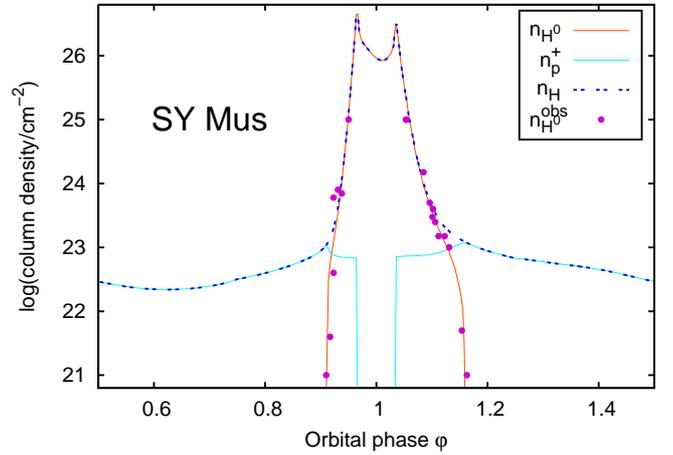}}
\end{center}
\caption[]{United column density model, $n_{\rm H}$ and its components,
$n_{\rm H^0}$ and $n^+_{\rm p}$;
 $n^{\rm obs}_{\rm H^0}$ are the measured column densities of the
neutral hydrogen from \cite{du99}. 
During the eclipse, column 
densities are computed from the observer to the surface of the RG.
          }
\label{nHvsetky}
\end{figure}

(iii)
We determined values of $n_{\rm H^0}(b)$ and 
$n^+_{\rm p}(b)$ functions for the unified 
model of the velocity profile and ionization structure 
according to Eqs.~\eqref{nHnum} and \eqref{nH+num}, using 
$R_{\rm g}$ as the length unit. Their sum provides the total hydrogen column density 
$n_{\rm H}$ (see Fig.~\ref{nHvsetky}).

%

\begin{figure*}
\centering
\begin{center}
\resizebox{16cm}{!}{\includegraphics[angle=0]{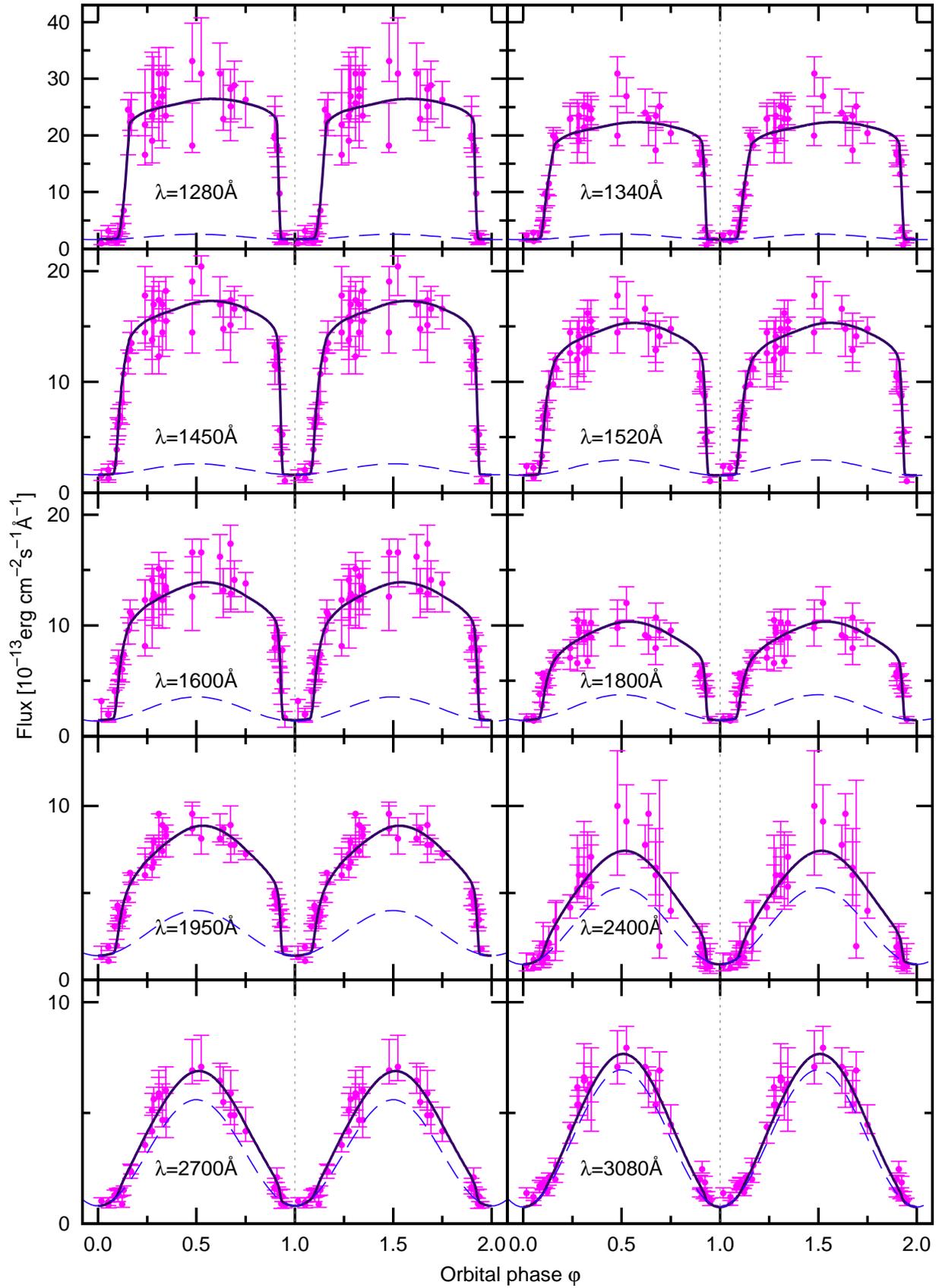}}
\end{center}
\caption[]{LCs of SY\,Mus (circles with errorbars) and their 
models (solid lines). Contribution from the nebula is depicted by dashed 
lines. Vertical dotted lines represent the position of the inferior 
conjunction of the giant.
          }
\label{10lc}
\end{figure*}

\begin{figure}
\centering
\begin{center}
\resizebox{\hsize}{!}{\includegraphics[angle=0]{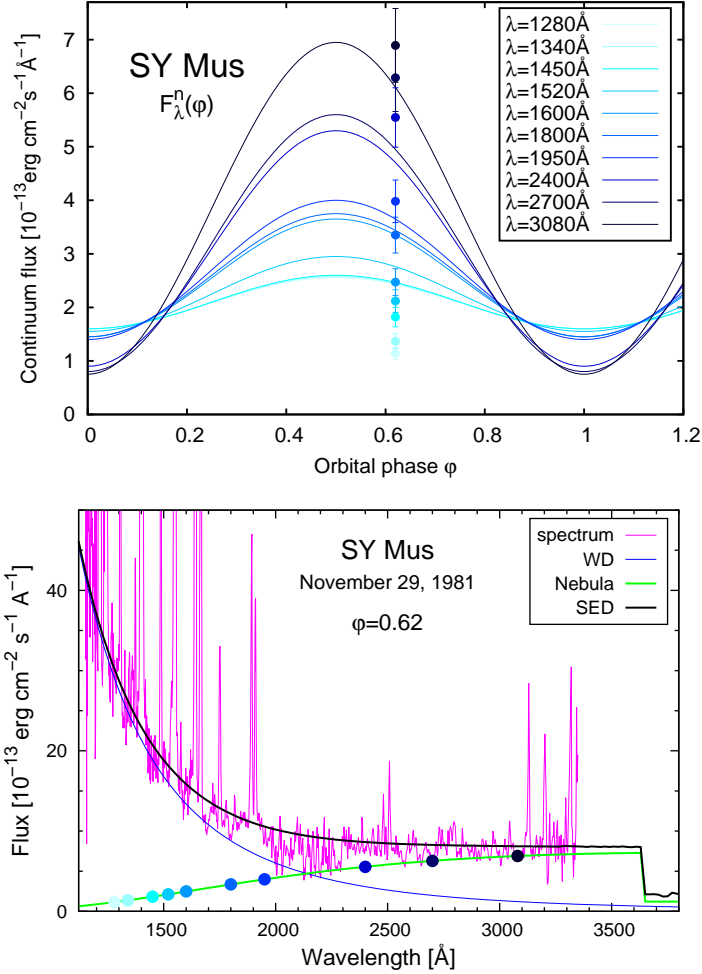}}
\resizebox{\hsize}{!}{\includegraphics[angle=-90]{sy_sed_062mod.eps}}
\end{center}
\caption[]{\textit{Top}: The model variations of the nebular continuum 
radiation in SY~Mus (Eq.~\ref{Fneb}) at 10 wavelengths (solid lines). 
They are compared with the nebular fluxes at $\varphi=0.62$ from the 
model SED (circles with errorbars). \textit{Bottom}: The model SED of 
SY Mus at $\varphi=0.62$. Circles at the nebular component correspond 
to those plotted at the top panel.
          }
\label{neb}
\end{figure}

\begin{figure}
\centering
\begin{center}
\resizebox{\hsize}{!}{\includegraphics[angle=0]{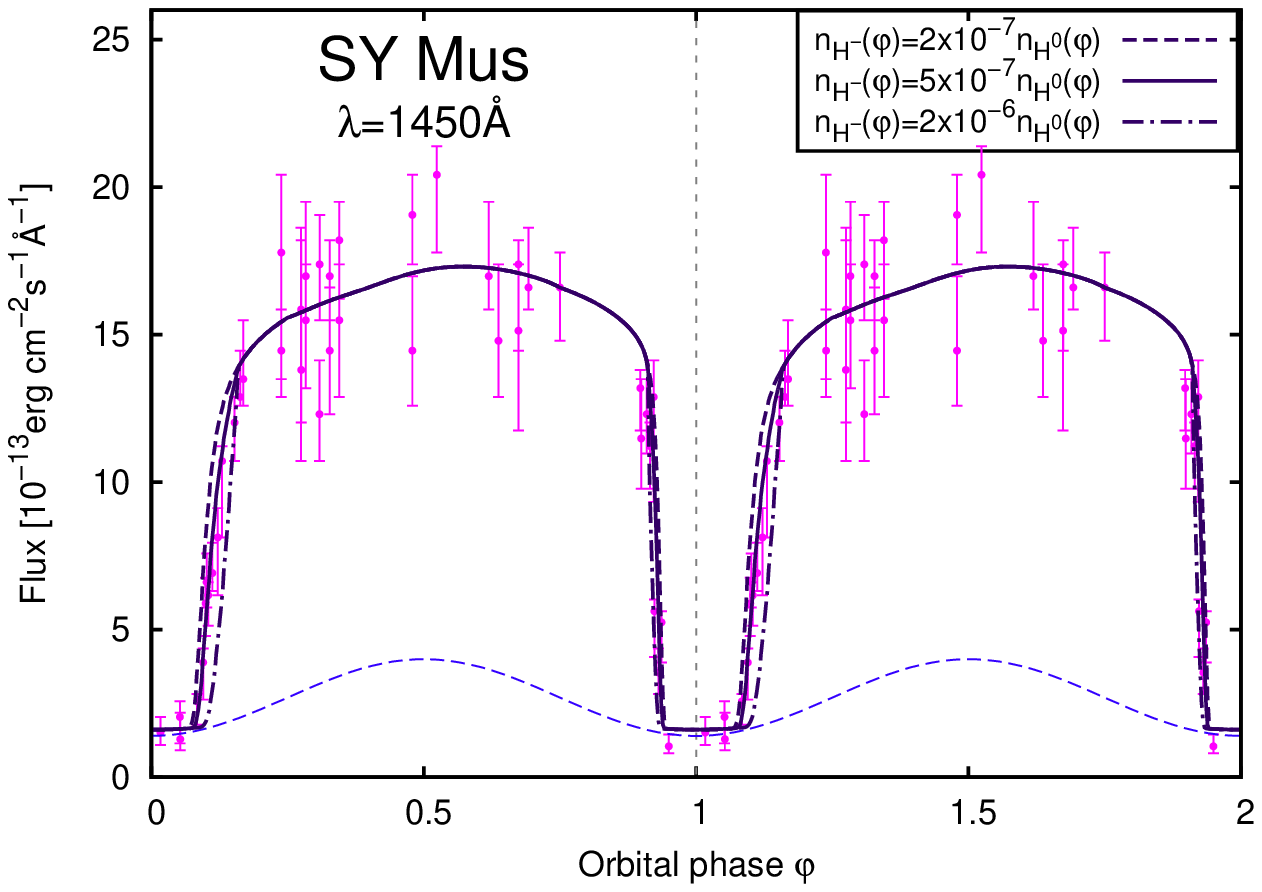}}
\vskip 4mm
\resizebox{\hsize}{!}{\includegraphics[angle=0]{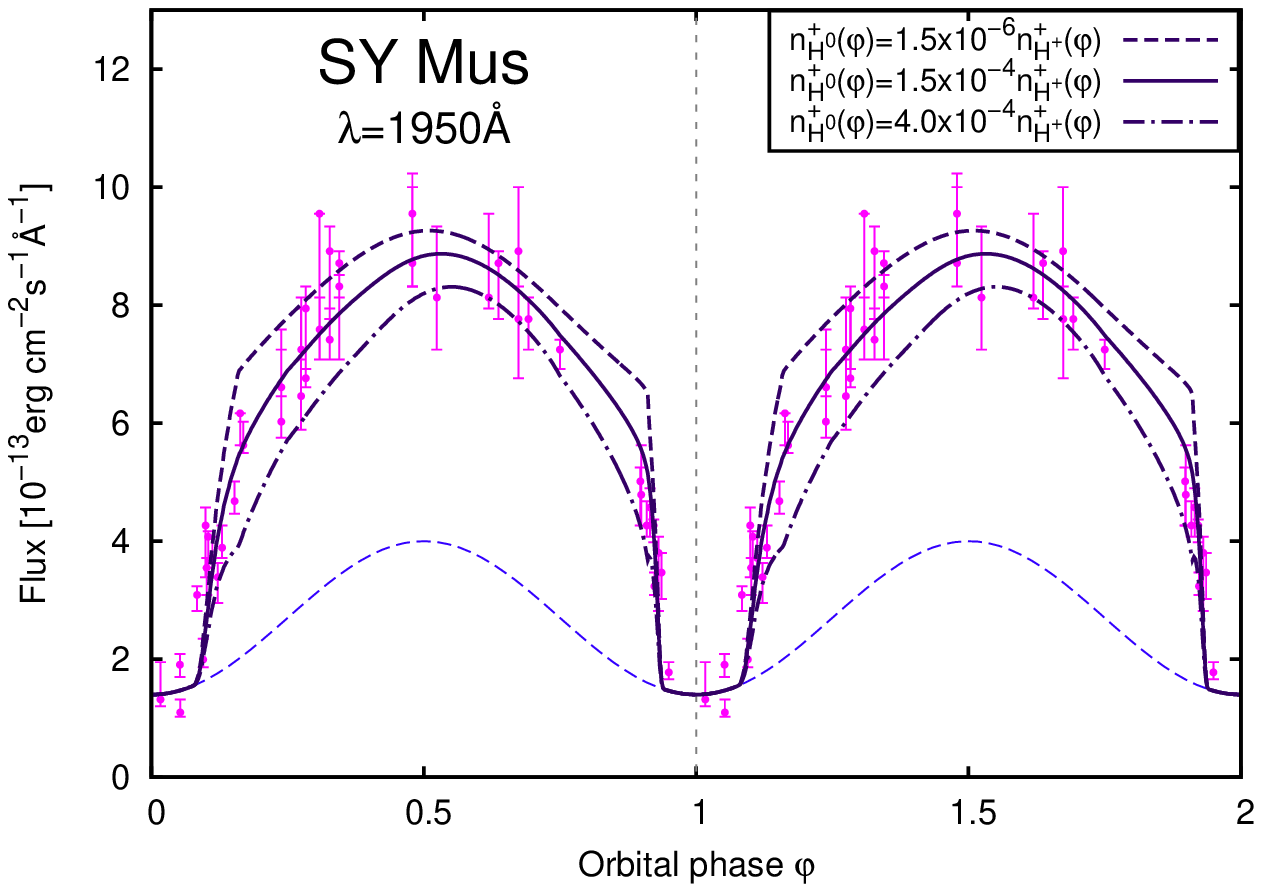}}
\end{center}
\caption[]{
Representative LC models for different column densities 
$n_{\rm H^-}(\varphi)$ (top) 
and $n^+_{\rm H^0}(\varphi)$ (bottom).
          }
\label{HmH0}
\end{figure}

(iv)
Finally, we modelled the continuum fluxes of SY~Mus
(Table~1) using Eq.~\eqref{Ftot}. 
The resulting models are depicted in Fig.~\ref{10lc}. Corresponding 
contributions from the nebula and a comparison with the values from the model 
SED at $\varphi=0.62$ is in Fig.~\ref{neb}. The model SED suggests a slightly 
more rapid increase of the continuum nebular flux with wavelength than the LC 
model. However, parameters of the nebular contribution are subject to variation 
from cycle to cycle \citep{sk05}.
The modelling yielded column densities
$n_{\rm H^-}(\varphi)=5.0\times 10^{-7}n_{\rm H^0}(\varphi)$  and  
$n^+_{\rm H^0}(\varphi)=1.5\times 10^{-4}n^+_{\rm p}(\varphi)$. We estimated their uncertainties by the LC models for $\lambda=1450$\,\AA\, and 
$\lambda=1950$\,\AA, respectively (see Fig.~\ref{HmH0}). 
Corresponding values of $n_{\rm H^-}$ and $n^+_{\rm H^0}$ 
are consistently used in all 10 LC models, from $\lambda=1280$\,\AA\, to 
$\lambda=3080$\,\AA.

%

\section{Discussion}
\label{dis}

Results of our analysis are limited by simplifying 
assumptions and quality of the data. Basic assumption we made is that 
the wind from the giant
is composed solely by 
the hydrogen in different states/forms and that it flows radially. 
Further, we neglected an influence of the hot component wind,
because it is a factor of $\approx 10$ weaker than that from the RG 
during quiescent phases \citep[][]{sk06}. 
As the real composition and the flow of the wind is more complex, 
these assumptions represent 
sources of systematic errors.

Further simplification is given by the selection of sources 
attenuating the continuum radiation, 
i. e. H$^0$ atoms, H$^-$ ions and free 
electrons. We tested the sensitivity of our LC model on the 
value of product of the column density and the continuum cross-section 
of atom/ion X, $n_{\rm X}\sigma_{\rm X}$. We have found
no detectable attenuation effect to the LC for
$n_{\rm X}\sigma_{\rm X}/n^+_{\rm p}\lesssim 10^{-26}$\,cm$^2$ in the ionized 
region and for $n_{\rm X}\sigma_{\rm X}/n_{\rm H^0}\lesssim 10^{-29}$\,cm$^2$ 
in the neutral region. Using these relations, we checked, whether H$_2^+$ and H$^-$ ions as 
other possible sources of scattering the continuum can 
contribute to the LC attenuation in the ionized region. We found 
that none of them has a measurable effect for the column density 
values from the model for symbiotic stars of \cite{sc97} or even higher 
values of column density according to the model for planetary nebulae of 
\cite{ag04}. Also the 
concentrations of free electrons and protons in neutral area are too 
low \citep{sc97,cr06} to have a recognizable effect on the continuum UV 
radiation. However, due to the simplifications in our model, we cannot 
exclude the effect of other species than those included in our model and, 
therefore, we consider the resulting functions $n^+_{\rm H^0}(\varphi)$ 
and $n_{\rm H^-}(\varphi)$ as the upper limits estimate
of the column densities 
of H$^0$ atoms in the ionized region and H$^-$ ions in the neutral region. 

In determining column densities $n_{\rm H^-}(\varphi)$ 
and $n^+_{\rm H^0}(\varphi)$, we preliminary approximated these functions using expressions that 
qualitatively correspond to the $n_{\rm H^-}/n_{\rm H^0}$ and 
$n^+_{\rm H^0}/n^+_{\rm p}$ profiles of \cite{ag04} for planetary 
nebulae. However, these expressions and their  
modifications did not represent observed fluxes near to the eclipse. Therefore, 
H$^-$ and H$^0$ densities probably rapidly change at the ionization 
boundary, as in the models for H$^0$ and H$^+$ populations of \cite{sc97}. 
Thus, we simply used linear functions of the column density of the 
corresponding prevalent form of the hydrogen (Sect.~\ref{cur}, point (v)).
Therefore, functions $n_{\rm H^-}(\varphi)$, $n^+_{\rm H^0}(\varphi)$ and 
$n^+_{\rm e}(\varphi)$ depend on $n_{\rm H^0}(\varphi)$ or 
$n^+_{\rm p}(\varphi)$ that
are given by velocity profiles derived from observations.

A large scatter in the observed fluxes  
(Fig.~\ref{10lc}) does not allow to determine the
nebular continuum variations, $F^{\rm n}_\lambda(\varphi)$, accurately.
 To put a constraint on the resulting nebular component,
we required a monotone increasing 
of the nebular flux with wavelength for $\varphi\approx 0.5$ and its monotone 
decreasing for $\varphi\approx 1$, as in the models SED \citep{sk09}. Figure~\ref{neb} 
shows an example of the model SED at the orbital phase $\varphi=0.62$ and a 
comparison of its $F^{\rm n}_\lambda(0.62)$ fluxes with those from
our LC models. Both models are roughly consistent. However,
dependence 
of our model nebular flux on the wavelength is not monotone for 
$\varphi\approx 0.1$ and $0.9$ due to the worse quality of the data.

Further simplification is given by the approximation that the nebular continuum 
varies with the orbital phase as a sine function (Eq.~\eqref{Fneb}). Since during 
quiescent phases
the symbiotic nebula represents the ionized fraction of the wind from the giant,
the asymmetric wind distribution will produce a nebula, which is asymmetric with 
respect to the binary axis and thus its flux. However, with given uncertainties 
in the data, 
this effect is not recognizable even for LCs with dominant nebular
contribution, for $\lambda>2000$\,\AA\,(Fig.\ref{10lc}).
Comparing our resulting nebular fluxes with those of \cite{sk05} and 
\cite{sk09} from models SED at $\varphi=0.02$ and $\varphi=0.62$, we 
found different values by a factor $1.04 - 1.89$, which is given
by the scatter of the data at a given wavelength (also Fig.~\ref{neb}).

Furthermore, we considered a continuous change of
the velocity profile within the transition region (Eq.~\eqref{vtrans}),
which affects mainly the profiles around maxima and minima of
the model LCs. Applicability of our transition velocity profile is
supported by the resulting LC models that describe
the eclipse profiles of the observed LCs. Particularly, a good match 
was found for the LC at $\lambda=2700$\,\AA\, (the reduced $\chi$-squared 
$\chi^2_{\rm red}=0.4$).
Overall, the values of $\chi^2_{\rm red}$ of 
our models varies from the order of $0.1$ to the order of $10$. 

To improve the presented modelling, better quality spectra of the 
symbiotic systems are needed, preferably in such quantity that even single 
orbital cycle can be covered sufficiently, to avoid physical changes of the 
particular system with time. With this kind of data, refinement of the model 
parameters would be possible and also including the asymmetric nebula model 
would be meaningful.

An independent way how to probe the asymmetric distribution of 
the circumstellar matter in symbiotic binaries from observations, 
as suggested by our continuum analysis \citep[][ and this paper]{ss16}, 
is to investigate dramatic and strictly orbitally-related 
variation of the H$\alpha$ line profile observed in the spectra 
of other quiet symbiotic stars. The main goal is to model the 
orbital variations of both the absorption and emission component 
in the H$\alpha$ line profile as 
a function of the orbital phase, using our column density model. 
This means to investigate the absorbing/scattering layer of 
the neutral wind from above the RG photosphere and the line 
emitting H$^{+}$ region, with the aim to determine the H$\alpha$ 
line profile at each orbital phase. And in this way, to obtain the 
corresponding velocity profile of the circumstellar matter, its 
distribution with respect to the binary axis and the spherical 
equivalent of the mass-loss rate. 
Solving this task should also provide constraints for further 
theoretical modelling of the RG wind in symbiotic binaries, 
its transfer to and accretion onto the compact companion. 



%

\section{Conclusion}
\label{conc}

We modelled the IUE continuum fluxes of the eclipsing symbiotic system 
SY~Mus at 10 wavelengths from $1280$ to $3080$\,\AA\, (Table~1, available online). 
To evaluate attenuation of the WD radiation passing the wind from the
RG at different wavelengths and orbital phases, we considered
the Rayleigh scattering on neutral hydrogen atoms and the bound-free absorption
by negative hydrogen ion in the neutral wind region, and the Thomson scattering on 
free electrons and
the bound-free and free-free transitions on the neutral hydrogen atom in the ionized
region (Eqs.~\ref{tauN} and \ref{tauI}).

To obtain the column densities of relevant forms of hydrogen 
at all orbital phases, we used the 
asymmetric wind velocity profile (Sect.~\ref{vfi}) 
and the corresponding 
ionization boundary model (Fig.~\ref{ionhranSYMus}). 
Accordingly, we obtained the column density distribution of H$^0$ in the 
neutral region and H$^+$ in the ionized region (Fig.~\ref{nHvsetky}).
From the shape of the observed LCs, 
we obtained the distribution of the H$^-$ 
column density $n_{\rm H^-}(\varphi)/n_{\rm H^0}(\varphi)=5.0\times 10^{-7}$ 
that can be considered as the upper limit due to the simplifications 
in the model (Sect.~\ref{cur}, Fig.~\ref{HmH0}).
Resulting LC models of SY~Mus (Eq.~\eqref{Ftot}, 
Fig.~\ref{10lc}) match well the 
asymmetric shape of the observed flux profiles. The nebular contribution 
was approximated by a sine waves (Sect.~\ref{cur}, Fig.~\ref{neb}). 

This paper presents the first quantitative model of the UV LCs
of the S-type symbiotic star.
We found that the different shapes of descending and 
ascending parts of the observed UV continuum LCs of SY~Mus are
caused by the asymmetric distribution of the wind from the RG at the 
near-orbital region.

\begin{acknowledgements}
The authors thank the referee Hans Martin Schmid for constructive comments.
This article was created by the realisation of the project ITMS 
No.~26220120029, based on the supporting operational Research 
and development program financed from the European Regional 
Development Fund. 
This research was supported by a grant of the Slovak Academy of 
Sciences, VEGA No. 2/0008/17. 
This work was supported by the Slovak Research and Development
Agency under the contract No. APVV-15-0458.
\end{acknowledgements}


\begin{thebibliography}{}
\bibliographystyle{aa}

\bibitem[Aleman \& Gruenwald(2004)]{ag04}
         Aleman, I., \& Gruenwald, R. 2004, ApJ, 607, 865

\bibitem[Andronov(1986)]{an86}
         Andronov, I. L. 1986, Astron. Zh. 63, 274

\bibitem[Bell et al.(1984)]{be84}
         Bell, S. A., Hilditch, R. W., \& Hoyle, F. 1984, MNRAS, 208, 123
         
\bibitem[Bohm \& Aller(1947)]{ba47}
         Bohm, D., \& Aller, L. H. 1947, ApJ, 105, 131

\bibitem[Bopp \& Noah(1980)]{bn80}
         Bopp, B. W., \& Noah, P. W. 1980, Publications of the ASP, 92, 717
         
\bibitem[Boyarchuk(1967)]{bo67}
         Boyarchuk, A. A. 1967, SvA, 11, 8
         
\bibitem[Crowley(2006)]{cr06}
        Crowley, C., 2006, {\it Red Giant Mass-Loss: Studying 
        Evolved Stellar Winds with FUSE and HST/STIS}, Thesis, 
        Trinity College Dublin
         
\bibitem[Doughty \& Fraser(1966)]{df66}
         Doughty, N. A., \& Fraser, P. A. 1966, MNRAS, 132, 267         
         
\bibitem[Dumm et al.(1999)]{du99}
         Dumm, T., Schmutz, W., Schild, H., \& Nussbaumer, H. 
         1999, A\&A 349, 169
         
\bibitem[Elsner et al.(1980)]{el80}
         Elsner, R. F., Ghosh, P., Darbro, W., \& Weisskopf, M. C. 1980,
         ApJ, 239, 335         
         
\bibitem[Fern\' andez-Castro et al.(1988)]{fc88}
         Fern\' andez-Castro, T., Cassatella, A., Gim\' enez, A.,
         \& Viotti, R. 1988, ApJ, 324, 1016
         
\bibitem[Formiggini \& Leibowitz(1990)]{fl90}
         Formiggini, L., \& Leibowitz, E. M. 
         1990, A\&A, 227, 121
         
\bibitem[Geltman(1962)]{ge62}
         Geltman, S. 1962, ApJ, 136, 935         
         
\bibitem[Gray(2005)]{gr05}
        Gray, D. F., 2005, {\it The Observation and Analysis of Stellar 
        Photospheres}, Cambridge University Press
        
\bibitem[Greiner et al.(1997)]{gr97}
         Greiner, J., Bickert, K., Luthardt, R., et al. 1997, A\&A, 322, 576
        
         
         
\bibitem[Kato et al.(2012)]{ka12}
         Kato, M., Miko\l ajewska, J., \& Hachisu, I. 2012, Baltic 
         Astronomy, 21, 157     
         
\bibitem[Kato et al.(2013)]{ka13}
         Kato, M., Hachisu, I., \& Miko\l ajewska, J. 2013, ApJ, 763, 5 
         
\bibitem[Knill et al.(1993)]{kn93}
         Knill, O., Dgani, R., \& Vogel, M. 
         1993, A\&A, 274, 1002         
         
\bibitem[Kolotilov et al.(2002)]{ko02}
         Kolotilov, E. A., Tatarnikova, A. A., Shugarov, S. Yu., \& Yudin, 
         B. F. 2002, Astronomy Letters, Vol. 28, No. 9, 620          
         
\bibitem[M\"urset et al.(1991)]{mu91}
         M\"urset, U., Nussbaumer, H., Schmid, H. M., \& Vogel, M. 
         1991, A\&A 248, 458  
         
\bibitem[M\"urset \& Schmid(1999)]{ms99}
         M\"urset, U., \& Schmid, H. M. 1999, A\&AS, 137, 473 
         
\bibitem[Nussbaumer \& Vogel(1987)]{nv87}
         Nussbaumer, H., \& Vogel, M. 1987, A\&A, 182, 51
         
\bibitem[Nussbaumer et al.(1989)]{nu89}
         Nussbaumer, H., Schmid, H. M., \& Vogel, M. 
         1989, A\&A, 211, L27
         
\bibitem[Pereira et al.(1995)]{pe95}
         Pereira, C. B., Vogel, M. \& Nussbaumer, H. 
         1995, A\&A, 293, 783
         
\bibitem[Popper(1961)]{po61}
         Popper, D. M. 1961, ApJ 133, 148                           
         
\bibitem[Pribulla et al.(2011)]{pr11}
         Pribulla, T., Va\v nko, M., Chochol, D., Hamb\' alek, \v L., 
         \& Parimucha, \v S. 2011, AN, 332, No. 6, 607        
         
\bibitem[Proga et al.(1998)]{pr98}
         Proga, D., Kenyon, S. J. \& Raymond, J. C.
         1998, ApJ, 501, 339          

\bibitem[Pustylnik et al.(2007)]{pu07}
         Pustylnik, I., Kalv, P., Harvig V., \& Aas, T. 2007, A\&AT, 
         Vol. 26, Nos. 4-5, 339

\bibitem[Samec et al.(2009)]{sa09}
         Samec, R. G., Figg, E. R., Melton, R., et al. 
         2009,  in Kosovichev, A. G., 
         Andrei, A. H., Rozelot, J. P., eds Sollar and Stellar Variability:
         Impact on Earth and Planets, 
         Proceedings IAU Symposium 264, p. 75
         
\bibitem[Schmid(1995)]{sc95}
         Schmid, H. M. 1995, MNRAS, 275, 227         
         
\bibitem[Schmid et al.(1999)]{sc99}
         Schmid, H. M., Krautter, J., Appenzeller, I., et al. 
         1999, A\&A, 348, 950
         
\bibitem[Schwank et al.(1997)]{sc97}
         Schwank, M., Schmutz, W. \& Nussbaumer, H. 
         1997, A\&A, 319, 166
         
\bibitem[Seaquist et al.(1984)]{se84}
         Seaquist, E. R., Taylor, A. R., \& Button, S. 
         1984, ApJ, 284, 202 (STB)     
         
\bibitem[Seker\'{a}\v{s} \& Skopal(2012)]{ses12}
         Seker\'{a}\v{s}, M., \& Skopal, A. 2012, MNRAS, 427, 979               
         
\bibitem[Shagatova et al.(2016)]{ss16}
         Shagatova, N., Skopal, A., \& Carikov\'a, Z. 2016, A\&A, 588, A83          
         
         
\bibitem[Skopal(1998)]{sk98}
         Skopal, A. 1998, A\&A, 338, 599    
         
\bibitem[Skopal(2001)]{sk01}
         Skopal, A. 2001, A\&A, 366, 157 
         
\bibitem[Skopal(2005)]{sk05} 
         Skopal, A. 2005, A\&A, 440, 995  
         
\bibitem[Skopal(2006)]{sk06} 
         Skopal, A. 2006, A\&A, 457, 1003  
         
\bibitem[Skopal(2009)]{sk09}
         Skopal, A. 2009, New Astronomy, 14, 336                                                       
         
                                
\bibitem[Skopal et al.(2002)]{sk02}
         Skopal, A., Va\v{n}ko, M., Pribulla, T., et al. 2002, CoSka, 32, 62            
                                           
\bibitem[Skopal \& Shagatova(2012)]{ss12}
         Skopal, A., \& Shagatova, N. 2012, A\&A, 547, A45  
         
\bibitem[Skopal et al.(2012)]{ss+12}
         Skopal, A., Shugarov, S., Va\v nko, M., et al. 2012, AN, 333, 
         No. 3, 242        
         
         
\bibitem[Watarai et al.(2005)]{wa05}
         Watarai, K., Takahashi, R., \& Fukue, J. 2005, PASJ, 57, 827   
         
\bibitem[Wi\k{e}cek et al.(2010)]{wi10}
         Wi\k{e}cek, M., Miko\l ajewski, M., Tomov, T., et al. 2010, eprint
         arXiv:1003.0608
       
\bibitem[Yuan(2010)]{yu10}
         Yuan, J. 2010, AJ 139, 1801

\bibitem[Zhou \& Leung(1990)]{zl90}
         Zhou, D.-Q., \& Leung, K.-Ch. 1990, ApJ, 355, 271
         
%
\end{thebibliography}
\end{document}